\documentclass[twocolumn]{aastex631}
\usepackage[left]{lineno}
\usepackage{epstopdf}
\usepackage{ulem}
\usepackage{amssymb}
\usepackage{natbib}
\usepackage{times}
\usepackage{graphicx}
\usepackage[usenames,dvipsnames]{pstricks}

\usepackage{psfrag}
\usepackage{lipsum}
\usepackage{epstopdf}

\shorttitle{GRB 230307A precursor}
\shortauthors{Dichiara et al.}

\begin{document}

\title[GRB 230307A precursor]{ 
A luminous precursor in the extremely bright GRB 230307A}

\author[0000-0001-6849-1270]{Dichiara S.}
\affiliation{Department of Astronomy and Astrophysics, The Pennsylvania State University, 525 Davey Lab, University Park, PA 16802, USA}

\author[0000-0002-1612-2585]{Tsang D.}
\affiliation{Department of Physics, University of Bath, Claverton Down, Bath, BA2 7AY, UK}

\author[0000-0002-1869-7817]{Troja E.}
\affiliation{Department of Physics, University of Rome - Tor Vergata, via della Ricerca Scientifica 1, 00100 Rome, IT}

\author{Neill D.}
\affiliation{Department of Physics, University of Bath, Claverton Down, Bath, BA2 7AY, UK}

\author{Norris J. P.}
\affiliation{Department of Physics, Boise State University, Boise, ID, USA}

\author[0000-0003-0691-6688]{Yang Y.-H.}
\affiliation{Department of Physics, University of Rome - Tor Vergata, via della Ricerca Scientifica 1, 00100 Rome, IT}


\begin{abstract}


GRB 230307A is an extremely bright long duration GRB with an observed gamma-ray fluence of $\gtrsim$3$\times$10$^{-3}$ erg cm$^{-2}$ (10--1000 keV), second only to GRB 221009A.
Despite its long duration, it is possibly associated with a kilonova, thus resembling the case of GRB 211211A. 
In analogy with GRB 211211A, we distinguish three phases in the prompt 
gamma-ray emission of GRB 230307A: 
an initial short duration, spectrally soft emission; a main long duration, spectrally hard burst; a temporally extended and spectrally soft tail. 
We intepret the initial soft pulse as a bright precursor to the main burst and compare its properties with models of precursors from compact binary mergers.  We find that to explain the brightness of GRB 230307A, a magnetar-like ($\gtrsim 10^{15}$ G) magnetic field should be retained by the progenitor neutron star.
Alternatively, in the post-merger scenario, the luminous precursor could point to the formation of a rapidly rotating massive neutron star.

\end{abstract}



\section{Introduction}
\label{sec:introduction}

Gamma-ray bursts (GRBs) are divided into two main phenomenological classes \citep{Kouveliotou1993}: long-duration ($>$2 s) GRBs produced by the collapse of massive stars, and short-duration ($<$2 s) GRBs thought to be caused by the coalescence of two compact objects, such as neutron stars (NSs) or black holes (BHs). 
In recent years, new sub-classes of GRBs have also been identified, such as short GRBs with extended emission \citep{Norris06} and peculiar long GRBs with hybrid high-energy properties, such as GRB 060614 and GRB 211211A \citep{Gehrels06,Rastinejad2022,Troja22}. 
Their gamma-ray emission is characterized by a main spectrally-hard burst followed by a long-lasting tail of spectrally softer emission. Despite their long duration, these bursts are not followed by bright supernovae and thought to be produced by compact object mergers.

Recently, the extremely bright GRB 230307A was proposed as a possible member of this new GRB class, although its prompt gamma-ray phase does not strictly follow the typical phenomenology of these events \citep{Yuhan2023}. 
Its gamma-ray emission displays a soft-hard-soft spectral evolution, more typical of standard long GRBs. 
In this work, we explore the hypothesis that the early spectrally softer emission of GRB 230307A is instead a precursor, preceding the main bright burst composed by a hard peak and a soft tail.

\begin{figure*}[t]
\centering
\includegraphics[scale=0.48]{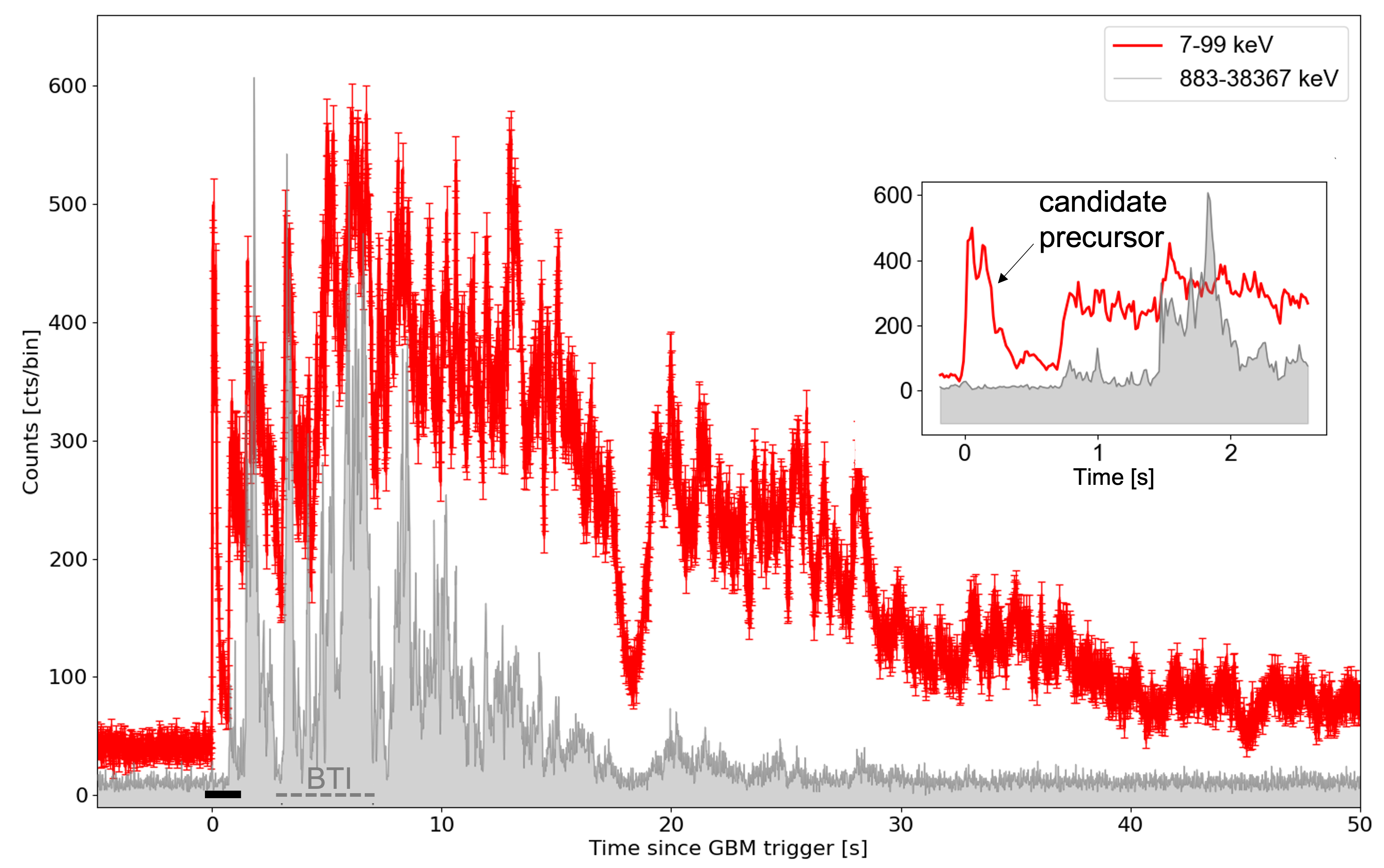}
 \caption{{\it Fermi}/GBM light curves of GRB 230307A in two energy bands, soft (7-99 keV) and hard (883-38367 keV). 
The time bin is 16 ms. 
Bad time intervals (BTIs) are indicated by a dashed horizontal line. 
The insert zooms in on the first 2.5~s (marked by the thick horizontal bar) to highlight the different spectral properties of the first peak, identified as a possible precursor.
}
 \label{fig:lcurves}
\end{figure*}

Numerous studies in the past have emphasized the presence of precursor signals in both long GRBs \citep{Burlon09,Lazzati05} and short GRBs \citep{Troja2010, Wang2020}. A precursor was also identified in the prompt emission of GRB 211211A \citep{Troja22, Xiao22}. 
Precursors exhibit lower luminosity and often a shorter duration than the main prompt episode \citep{Zhong2019,Li2021, Coppin2020}.
Several theories were proposed to explain the different characteristics associated with these precursors, encompassing spectral shape, duration, and quiescent times. 
A weak precursor with a quasi-thermal spectrum is a direct prediction of the standard fireball model \citep{MeszarosReees2000,Daigne2002}.
Other models are instead related to the nature of the central engine, with the precursor marking
the birth of a massive NS later collapsing into a BH \citep{WangMeszaros2007,Lipunov2008}. 
In other 
scenarios, the precursor is related to the strong interaction between the two compact objects right before they merge. 
Luminous high-energy emission is predicted by models that involve the shattering of the NS crust during the inspiral phase \citep[e.g., ][]{Tsang2012,Palenzuela13,Suvorov2020, Neill2021}
or the magnetospheric interactions between the neutron star and another compact object \citep[e.g., ][]{McWilliams2011,Most2020,Beloborodov2021,Cooper2023}.
\citealt{Schnittman2018} suggested that, 
regardless of their power source, precursors produced in the pre-merger phase should display a modulated temporal profile induced by the stars' orbital motion. 
While, in general, precursors associated to short GRBs or short GRBs with extended emission are few and faint, the extreme brightness of GRB 230307A allow us to put these models to test. 

The paper is organized as follows: in Section~\ref{sec:observations} we describe the data reduction process together with the timing and spectral analysis of GRB 230307A. In Section~\ref{sec:discussion} we discuss the results, illustrating the different models proposed to explain the nature of the precursor emission and highlighting the possible implications. In Section~\ref{sec:summary} we summarize our results and conclusions.

Uncertainties are quoted at the 1-$\sigma$ confidence level and upper limits are given at a 2-$\sigma$ level throughout the paper, unless stated otherwise. Standard $\Lambda$CDM cosmology \citep{Planck2018} was adopted. \\

\section{Observations and Data Analysis}
\label{sec:observations}

On March 7, 2023 at 15:44:06.67 UT, hereafter referred as a T$_0$, the {\it Fermi}/GBM triggered on GRB 230307A \citep{2023GCN.33405....1F}.
The burst was also observed by several other missions including GECAM \citep{GCN33406}, Konus/WIND \citep{GCN33427}, AGILE \citep{GCN33412} and the Solar Orbiter STIX \citep{GCN33410}. These detections confirmed the extremely bright nature of the transient event and enabled the Interplanetary Gamma-Ray Burst Timing Network (IPN) to provide a refined localization \citep{GCN33413}. 
A preliminary estimate of its broadband (10-1000 keV) fluence reached a remarkable value of approximately 3$\times$10$^{-3}$ erg cm$^{-2}$ \citep{GCN33411,GCN33427}, marking the second highest value ever recorded \citep{Burns2023GCN, Oconnor2023}.

To conduct our analysis, we use {\it Fermi}/GBM’s time-tagged event (TTE) data obtained from three NaI detectors (N6, N8, and Na) and the BGO detector B1. 
We processed the data using the HEASoft 
\citep[version 6.30.1;][]{2014ascl.soft08004N} and the Fermitools 
software packages \citep[version 2.0.8;][]{2019ascl.soft05011F} following standard procedures \footnote{https://fermi.gsfc.nasa.gov/ssc/data/p7rep/analysis/scitools/gbm\_grb\_analysis.html}.  
The spectral analysis was conducted using RMFIT v.4.3.2 \citep{2014ascl.soft09011G}. 
Due to the extremely high flux of GRB 230307A, the data suffers from losses caused by electronic bandwidth limits \citep{GCN33551}. As a result, certain time intervals are affected by pile-up effects \citep[from $T_0 + $3 s to $T_0 + $7 s]{GCN33551} and were excluded from the analysis.

\subsection{Temporal analysis}
\label{sec:temp}
Figure~\ref{fig:lcurves} shows the GRB light curves  extracted with a time bin resolution of 16 ms. We compare the temporal profiles in two energy bands, a soft energy band ($7-99$ keV) and  a hard energy band ($>800$ keV). 
From Figure~\ref{fig:lcurves} we note that most of the emission in the hard energy band is concentrated within the first 18 s, whereas the longer tail of emission, extending up to $\sim$45 s, consists of softer energy photons. 

Using the standard  BATSE energy range (50--300 keV), we derive a duration $T_{90}$ of $\sim$33 s.
This value is defined as the time interval over which the cumulative number counts increase from 5\% to 95\% above background \citep{Kouveliotou1993}, and may thus be slightly overestimated due to pile-up effects. However, even assuming that the count rate is twice the observed value during the bad time interval, the resulting
$T_{90}$ would shorten by only 3~s and GRB 230307A would still be classified as a long GRB. 
In this respect,  it differs from the standard case of short GRBs with extended emission, whose spectrally hard pulse lasts less than 2 s, and resembles the long GRB 211211A.  

The inset of Figure~\ref{fig:lcurves} zooms in the first 2.5 sec of emission. The first bright pulse that triggered {\it Fermi}/GBM is visible only in the soft energy band, whereas the spectrally hard ($>$800 keV) emission does not start until $\approx T_0 +$ 1 s. 
The first pulse is characterized by a short duration, lasting from $T_{0}$ to $T_{0}+$0.4 s, and a double peaked structure. The two peaks are separated by $\sim$0.1 s. The gamma-ray flux decreases close to the background level after this first short signal and it increases again at $\sim T_{0}$+0.8 s, when the main part of the emission starts. 

The short duration of this initial pulse combined with the low flux (in comparison with the extremely bright main emission), the softer spectrum and the delay with respect to the onset of the main emission ($\sim$0.4 s) are distinctive features typical of GRB precursors. 
Previous studies of bright long GRBs show a possible relation between the duration of the quiescent time and that of the subsequent emission episode  \citep{Ramirez-Ruiz2001}. 
Neither GRB 230307A nor GRB 211211A follow this relation for long GRBs, supporting their classification as peculiar GRBs
\citep{Yuhan2023,Troja22,Sun2023}.
The short delay time of GRB 230307A ($\sim$0.4 s) is in the typical range of  other short GRB precursors detected by {\it Fermi} \citep[e.g. $<$1.7 s][]{Coppin2020} and its flux follow the observed relation between the precursor and the main emission \citep{Zhong2019}.

Two quantities commonly used for GRB studies are the spectral lag and minimum time scale variability. 
To derive the spectral lag we used the light curve extracted in the two standard energy ranges: 25-50 keV and 100-300 keV. 
At first we focused our timing analysis only on the precursor event (from $T_{0}$+0 sto $T_{0}$+0.18 s).
Using the same approach described in \citep{Norris06} and a 4-ms binned light curve we derived a spectral lag of 1.5$\pm$3.0 ms in this initial interval. 
To investigate possible evolution of the lag over the main emission we also measured this valued in different interval: from $T_{0}$+1.4 to $T_{0}$+2.9 s, from $T_{0}$+7.9 to $T_{0}$+13.5 s and from $T_{0}$+19.0 to $T_{0}$+23.4 s. We found -6.6$^{+9.1}_{-11.4}$ ms, 3.3$^{+5.2}_{-5.0}$ ms and 2.8$\pm$4.5 ms, respectively.
Intervals were selected in such a way that the count rates at the beginning and end points are almost the same. This deliberate choice aimed to enhance the efficacy of cross-correlation analysis and produce robust lag results.
We used the 8-ms binned light curve to study the main part and the extended tail of the prompt emission.
This was done in order to ensure the minimum number of counts per bin needed to perform a sensitive fits of the cross correlation function between the two channels. 

These spectral lags are consistent with 0 and are similar to the ones observed for short GRBs and short GRBs with extended emission \citep{Norris06}.

\begin{deluxetable}{ccc}[h!]
\tablecaption{Minimum variability time scale \label{tab:mv}} 
\tablehead{\colhead{Time Interval} & \multicolumn{2}{c}{Minimum variability} \\
s  & s & s \\
 & 8-1000 keV & 15-350 keV}
\startdata
$T_{0}$-0.064--$T_{0}$+0.32 &  0.017 $\pm$ 0.002 & 0.014 $\pm$ 0.002 \\
$T_{0}$+0.320--$T_{0}$+3    &  $<$0.012          & 0.010 $\pm$ 0.002 \\
$T_{0}$+7--$T_{0}$+18       &  0.015 $\pm$ 0.002 & 0.017 $\pm$ 0.002 \\
$T_{0}$+18--$T_{0}$+40      &  0.062 $\pm$ 0.004 & 0.063 $\pm$ 0.005 \\
\enddata
\tablecomments{All the reported times are observer frame}
\end{deluxetable}

\begin{figure}[h!]
\centering
\includegraphics[scale=0.4]{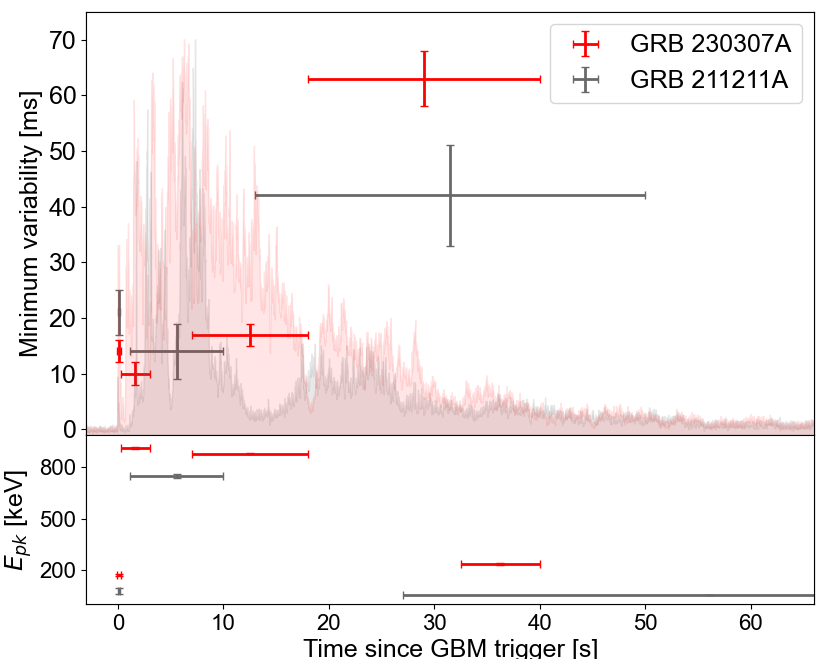}
 \caption{Minimum variability timescale (top) and spectral peak energy (bottom) as a function of time. Red and gray crosses mark the evolution measured for GRB 230307A and 211211A, respectively. The shaded re-normalized light curves (8--1000 keV) are GRB 230307A (red) and GRB 211211A 
 \citep[gray, from][]{Troja22}.
}
 \label{fig:mtv}
\end{figure}

We then derived the minimum time scale variability using the method described in \cite{Golkhou2014}. The value was calculated for different time intervals, representative of the different partes of the prompt emission: the precursor (-0.064 s - 0.32 s), the main hard peak (0.32-18 s, excluding the bad time interval 3-7 s), and the extended tail (18 - 40 s). 
Two energy bands were considered: the standard GBM broadband (8-1000 keV) and the standard {\it Swift} band (15-350keV) to allow for direct comparison with GRB 211211A and the rest of the {\it Swift} sample. 
Results are reported in Table~\ref{tab:mv} and Figure~\ref{fig:mtv}. 

During the precursor episode the minimum variability is 17$\pm$2 ms (see Figure~\ref{fig:mtv}).
As the prompt emission evolves, this variability changes. A fast variability ($<$12 ms) characterizes the signal between $T_{0}$+0.4 s and $T_{0}$+3 s (after the first peak and before the start of the bad time interval). The minimum variability timescale stays below $\sim$20 ms during the main emission 
(before $T_{0}$+18 s) and it increases up to $\sim$60 ms during the tail (see Table~\ref{tab:mv}).
Our values are slightly higher that the one reported by \cite{Camisasca2023GCN}, which was derived using a different approach.

It is interesting to note that the minimum variability evolves as observed for GRB 211211A (see Figure~\ref{fig:mtv}).
It is also worth to note that while the main signal seems to follow the typical hard-to-soft trend observed for other long GRBs commonly associated with a short-to-long variability, the precursor appears to have a peculiar short variability and a soft spectra (see the next Section~\ref{sec:spec}). This distinct behaviour deviates from the general trend observed between the pulse width and the energy \citep{Fenimore1995} and it is indicative of a different physical process associated with the precursor.

\subsection{Spectral analysis}
\label{sec:spec}
\begin{figure}[t]
\centering
\includegraphics[scale=0.37]{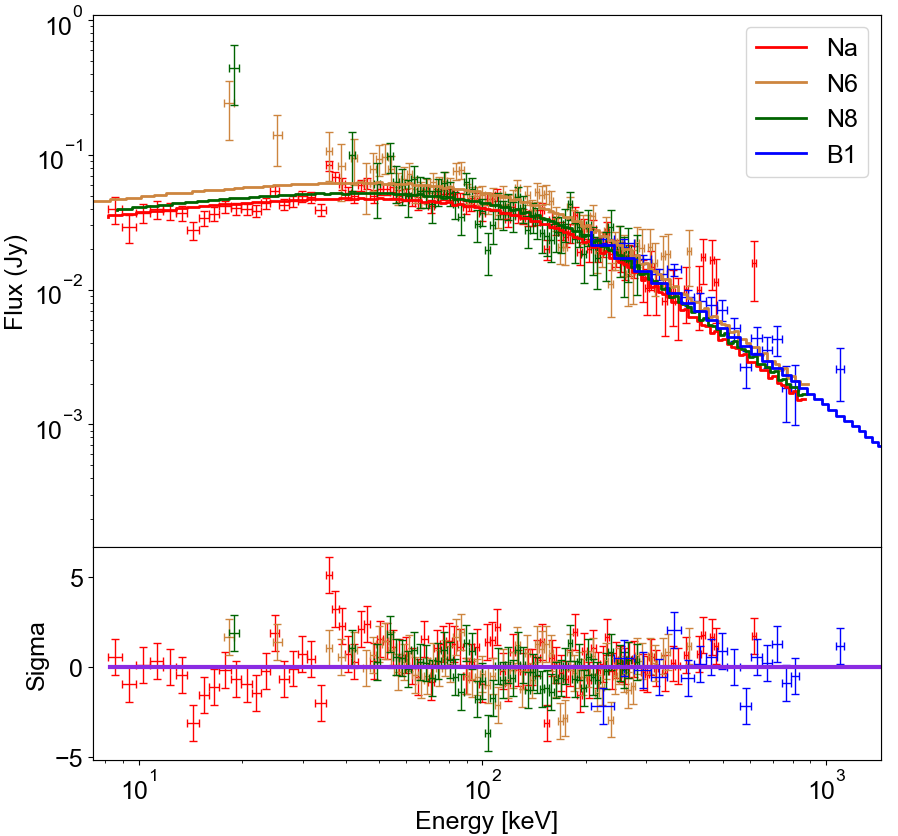}
 \caption{Spectral fit of GBM data integrated over the precursor (from T$_{0}$-0.064 to T$_{0}$+0.320) using the 3 most illuminated NaI detectors (N6, N8 and N10) and the BGO (B1). The inset text shows the best fit parameters obtained assuming a Band function.
}
 \label{fig:bandfit}
\end{figure}

The precursor spectra was obtained integrating the signal between $T_{0}$-0.064 s and $T_{0}$+0.320 s. The background was derived using the time interval that goes from $T_{0}$-89 s to $T_{0}$-7 s before the burst trigger and from $T_{0}$+162 s to $T_{0}$+273 s after the end of the prompt emission. The background light curve was modeled using a third degree polynomial. The spectral best fit results are presented in Table~\ref{tab:spectra} and in the bottom panel of Figure~\ref{fig:mtv}. The spectral models used for the analysis are: a simple black body (BB), Comptonized model \cite[e.g. a power-law with an exponential cut-off, as described in][]{Gruber2014}, a Band function \cite{Band1993} and a combination between a Band function and a BB. The best fit is obtained by a simple Band function (see Figure~\ref{fig:bandfit}). The combined Band+BB model provide a slightly better fit although the improvements is not significant enough to justify the presence of the thermal component. The average flux measured during the precursor episode is 3.3$\times$10$^{-5}$ erg cm$^{-2}$ s$^{-1}$ (10--1000 keV).

Through comprehensive multi-wavelength analysis of the GRB and its counterpart, a likely association with a nearby galaxy at z$\sim$0.065 has been established. These observations have provided compelling evidence, revealing the presence of a lanthanide-rich kilonova and effectively ruling out the possibility of a collapsar or a high-redshift origin \citep{Yuhan2023}.
Assuming that this nearby galaxy is the  actual host, we derive an average luminosity of $\sim$3.6$\times$10$^{50}$ erg s$^{-1}$.

In order to study the evolution of the spectra we fitted the data over different time intervals before $T_{0}$+3 s and after $T_{0}$+7 s. As discussed in the previous Section, the variability time scale is related with the spectral hardness (with shorter variability associated to harder spectra) except for the precursor that has a soft spectra and an surprisingly short variability.


\begin{deluxetable*}{cccccccc}
\tablecaption{Best fit Spectral parameters \label{tab:spectra}}
\tablehead{\colhead{Time Interval} & \colhead{Model} & \colhead{Mean Flux (10--1000 keV)} & \colhead{$E_{pk}$} & \colhead{$\alpha$} & \colhead{$\beta$} & \colhead{BB Temp} &   \colhead{C-STAT/}\\
s & & erg cm$^{-2}$ s$^{-1}$ &  keV & index & index & keV &  d.o.f}
\startdata
\hline
			 \multicolumn{8}{c}{Precursor}\\ \hline
$T_{0}$-0.064--$T_{0}$+0.32 & Black Body & 2.899($\pm$0.038)$\times$10$^{-5}$ & ---  & ---  & --- & $36.0 \pm 0.2$ & 3527.3/482 \\
$T_{0}$-0.064--$T_{0}$+0.32 &  Comptonized & 3.288($\pm$0.038)$\times$10$^{-5}$ &  $198.7 \pm 3.5$ & $-0.80 \pm 0.03$  & --- & --- & 606.1/481 \\
$T_{0}$-0.064--$T_{0}$+0.32 & Band        & 3.324($\pm$0.038)$\times$10$^{-5}$ &  $170.3 \pm 4.7$ & $-0.63 \pm 0.04$  & $-2.95 \pm 0.09$ & --- & 563.1/480 \\
$T_{0}$-0.064--$T_{0}$+0.32 & Band+BB     & 3.333($\pm$0.038)$\times$10$^{-5}$ &  $203.9 \pm 8.9$ & $-0.59 \pm 0.08$  & $-3.20 \pm 0.16$ & $15.3 \pm 1.4$ & 536.1/478 \\
\hline
			 \multicolumn{8}{c}{Main Emission}\\ \hline
$T_{0}$+0.320--$T_{0}$+3    & Band        & 1.196($\pm$0.003)$\times$10$^{-4}$ &  $914.6 \pm 5.4$ & $-0.47 \pm 0.01$  & $-5.5 \pm 0.3$ & --- & 954.4/485 \\
$T_{0}$+7--$T_{0}$+9       & Band        & 1.748($\pm$0.004)$\times$10$^{-4}$ &  $968.0 \pm 6.6$ & $-0.71 \pm 0.01$  & $-4.5 \pm 0.1$ & --- & 928.5/485 \\
$T_{0}$+9--$T_{0}$+11       & Band        & 1.577($\pm$0.004)$\times$10$^{-4}$ &  $946.8 \pm 6.7$ & $-0.73 \pm 0.01$  & $-4.8 \pm 0.2$ & --- & 1024.2/485 \\
$T_{0}$+11--$T_{0}$+13       & Band        & 1.170($\pm$0.003)$\times$10$^{-4}$ &  $804.5 \pm 6.8$ & $-0.90 \pm 0.01$  & $-5.0 \pm 0.4$ & --- & 1225.3/485 \\
$T_{0}$+13--$T_{0}$+15       & Band        & 9.533($\pm$0.031)$\times$10$^{-5}$ &  $781.7 \pm 8.4$ & $-1.07 \pm 0.01$  & $-4.8 \pm 0.4$ & --- & 1202.0/485 \\
$T_{0}$+15--$T_{0}$+18.5       & Band        & 4.916($\pm$0.016)$\times$10$^{-5}$ &  $634.6 \pm 8.0$ & $-1.17 \pm 0.01$  & $-4.5 \pm 0.4$ & --- & 1185.2/485 \\
$T_{0}$+18.5--$T_{0}$+22       & Band        & 4.592($\pm$0.016)$\times$10$^{-5}$ &  $617.2 \pm 8.3$ & $-1.23 \pm 0.01$  & $-5.0 \pm 0.9$ & --- & 1260.2/485 \\
$T_{0}$+22--$T_{0}$+25.5       & Band        & 3.594($\pm$0.014)$\times$10$^{-5}$ &  $468.4 \pm 6.8$ & $-1.27 \pm 0.01$  & $-5.1 \pm 1.5$ & --- & 1008.7/485 \\
$T_{0}$+25.5--$T_{0}$+29       & Band        & 2.457($\pm$0.012)$\times$10$^{-5}$ &  $392.6 \pm 9.1$ & $-1.50 \pm 0.01$  & $-4.7 \pm 1.8$ & --- & 967.30/485 \\
$T_{0}$+29--$T_{0}$+32.5       & Band        & 1.186($\pm$0.009)$\times$10$^{-5}$ &  $319.2 \pm 10.4$ & $-1.47 \pm 0.01$  & $-5.2 \pm 7.4$ & --- & 704.8/485 \\
$T_{0}$+32.5--$T_{0}$+40      & Band        & 9.870($\pm$0.056)$\times$10$^{-6}$ &  $233.0 \pm 4.8$ & $-1.50 \pm 0.01$  & $-4.7 \pm 2.9$ & --- & 957.5/485 \\
\enddata
\tablecomments{Best fit results obtained integrating the spectra over different time intervals. We used different models to fit the precursor spectra.}
\end{deluxetable*}

\section{Modeling of the precursor}
\label{sec:discussion}
Models for precursor emission in compact binary mergers can be grouped into two classes: pre-merger models, related to processes occurring before the merger, and post-merger models, which describe the 
 evolution of the remnant after the merger. 
In the pre-merger phase, the energy to power the precursor is extracted from the orbit, either through the NS magnetic field or from its crust.
In the post-merger phase, the energy is instead supplied by the central engine, thus allowing for a broader range of precursor luminosities and timescales. 

If GRB 230307A is truly associated with a kilonova at 291 Mpc -- and thus is a product of a merger involving at least one neutron star --  the extreme energetics of its first pulse ($L_{\rm iso} \sim 3.6 \times 10^{50}$ erg s$^{-1}$) challenge most models for precursor flares. In the following, we discuss possible scenarios for GRB 230307A.

\subsection{Pre-merger models: a Resonant Shattering Flare}

Among the pre-merger models, the resonant shattering flare model \citep{Neill2022, Tsang2012} can 
reproduce high luminosity precursors by extending slightly beyond the fiducial values presented in \citet{Neill2022, Tsang2012}, and requiring a significant surface magnetic field for one of the progenitors. Here, the crust-core interface mode ($i$-mode) of one of the merger progenitors is excited by tidal resonance. This mode is excited past the material breaking strain of the NS crust, causing the crust to fracture and shatter. Seismic oscillations in the NS can then drive perturbations of the surface magnetic field, which, if strong enough, can spark pair-photon fireball shells during the resonance window. Collisions between these shells lead to non-thermal gamma-ray emission.

\begin{figure}
\centering
\includegraphics[width=1.0\columnwidth,angle=0]{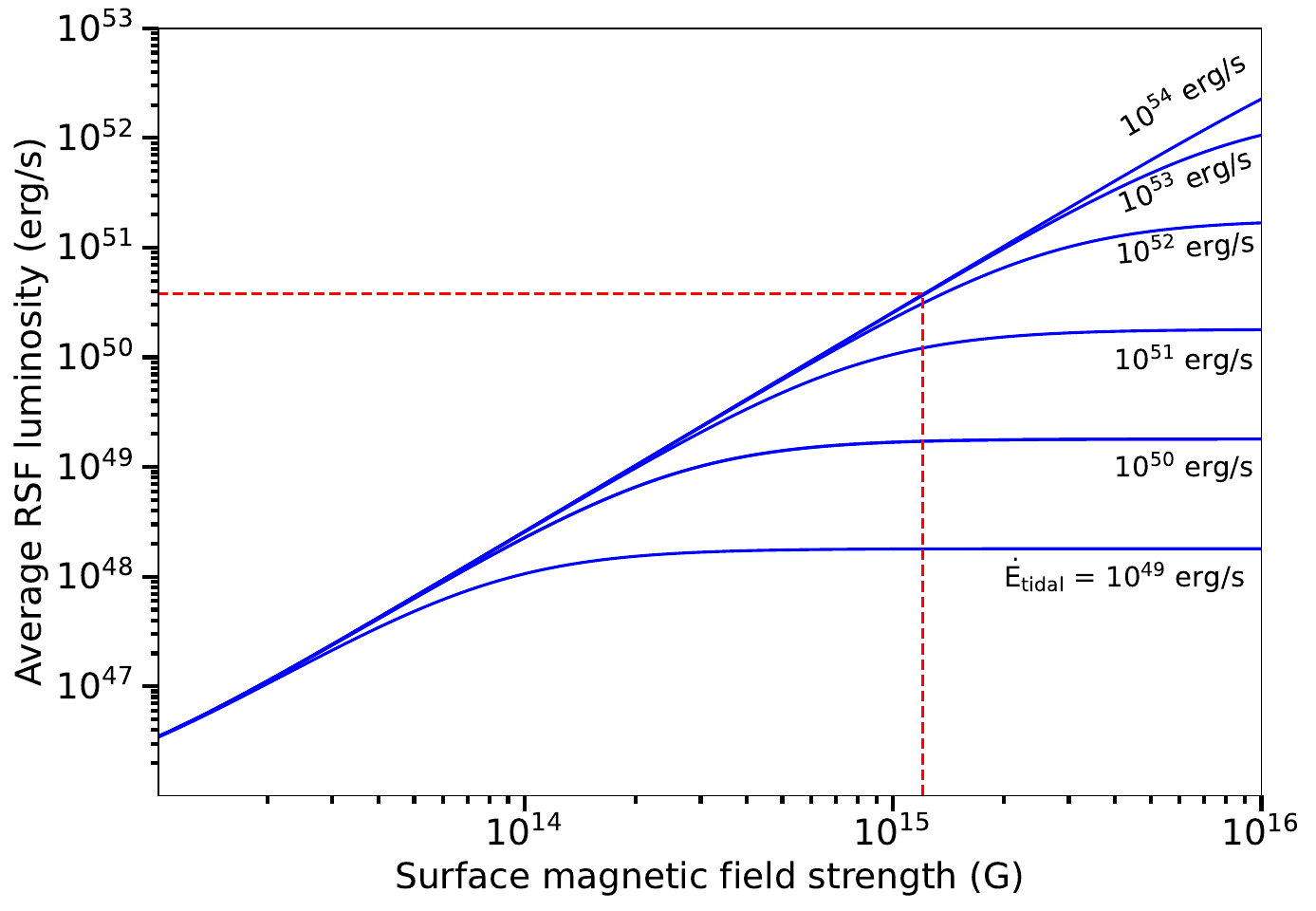}
\caption{The relationship between the average luminosity of a RSF and the strength of the magnetic field of the NS which produced it, assuming a $20\%$ gamma-ray efficiency. The relationship is shown for several different values for the rate at which energy is transferred to the $i$-mode from the binary orbit (equation~\ref{eq:Edot_tidal}). Red dashed lines indicate the luminosity of GRB 230307A's precursor, and the minimum magnetic field strength required to produce it.}
\label{BvsLFig}
\end{figure}

The total energy available in the Resonant Shattering flares (RSF) model is set by the tidal energy transfer rate given by \citep{Tsang2012}
\begin{equation}
\dot{E}_{\rm tidal} \simeq \left(\frac{3\pi}{40} \right)^{1/2} (2 \pi f_{\rm i-mode})^2 M_{\rm NS}^{1/2} R_{\rm NS} Q E_b^{1/2} \frac{q}{q+1}.
\label{eq:Edot_tidal}
\end{equation}
where $f_{\rm i-mode}$ is the $i$-mode frequency, $Q$ is the tidal overlap parameter, representing the coupling of the $i$-mode with the tidal field, $q$ is the binary mass ratio, and $E_b$ is the mode energy at which the breaking strain is reached. For fiducial values provided in \citet{Neill2021, Tsang2012} this can reach up to $\sim 10^{50}$ erg/s, which cannot produce the observed precursor with reasonable gamma-ray efficiency (see Figure~\ref{BvsLFig} which assumes a 20\% gamma-ray efficiency). This value could potentially be increased by an order of magnitude if the neutron star crust breaking strain and tidal overlap integral are a factor of a few larger than the conservative fiducial values assumed. 

Additionally, the surface magnetic field plays a significant role in limiting the luminosity; the magnetic field must be strong enough to extract the seismic energy from the crust, otherwise this energy remains trapped within the neutron star \citep{Tsang2012, Tsang2013}. If the precursor is indeed from an RSF event, this puts a lower limit on the surface magnetic field of $\gtrsim 1.2 \times 10^{15}$ G (see Figure \ref{BvsLFig}). This requires that a significant surface field remain present during the binary lifetime, which is unlikely if the field is confined to the crust, where Hall evolution and Ohmic dissipation would cause the field to decay in $\lesssim 10^6$ yr \citep{Gourgouliatos2016}. Instead, a field partially frozen into a superconducting core could provide support for a sufficiently long-lived surface field \citep[see e.g][]{Ho2017}.

In the simplified colliding fireball shell model, the spectral hardness of the non-thermal gamma-ray emission depends on the Lorentz factor of the fireball shells. This is determined by the breaking strain of the neutron star crust, as well as mass-loading of the pair-fireball by material from the neutron star surface. The duration of a RSF is closely related to the duration over which the $i$-mode is in resonance \citep{Neill2022}. To calculate this we follow \citet{Lai1994} except that we substitute their use of the stationary phase approximation with solving the equations for the evolution of the binary separation ($D$) and the orbital angular frequency ($\Omega$) due to the gravitational wave emission of two point masses in a circular binary. Note that we neglect relativistic effects that may become important late in the binary inspiral and effects of tidal interactions on the binary orbit, which will likely act to accelerate the inspiral, and so our results serve as an upper bound on the duration of resonance (particularly for more massive binaries, where the inspiral ends at lower orbital frequencies and thus closer to resonance). From the binary evolution we then calculate the cumulative tidal-field-strength-weighted phase difference between the $i$-mode and the binary orbit:
\begin{equation}
\phi(t')=\int^{t'}_{t_0}\frac{1}{D(t)^3}\exp(-2i\Phi+i(2\pi f_{\rm i-mode})t)dt
\label{eq:phase_evolution}
\end{equation}
\noindent (where the orbital phase function $\Phi=\int\Omega dt$ when assuming the NS which produces the RSF has zero spin, and $t_0$ is a time long before the resonance), which is proportional to the amplitude of the $i$-mode's oscillations and shows a sharp increase over a short time period around the resonance: the ''resonance window''.

\begin{figure}
\centering
\includegraphics[width=\columnwidth,angle=0]{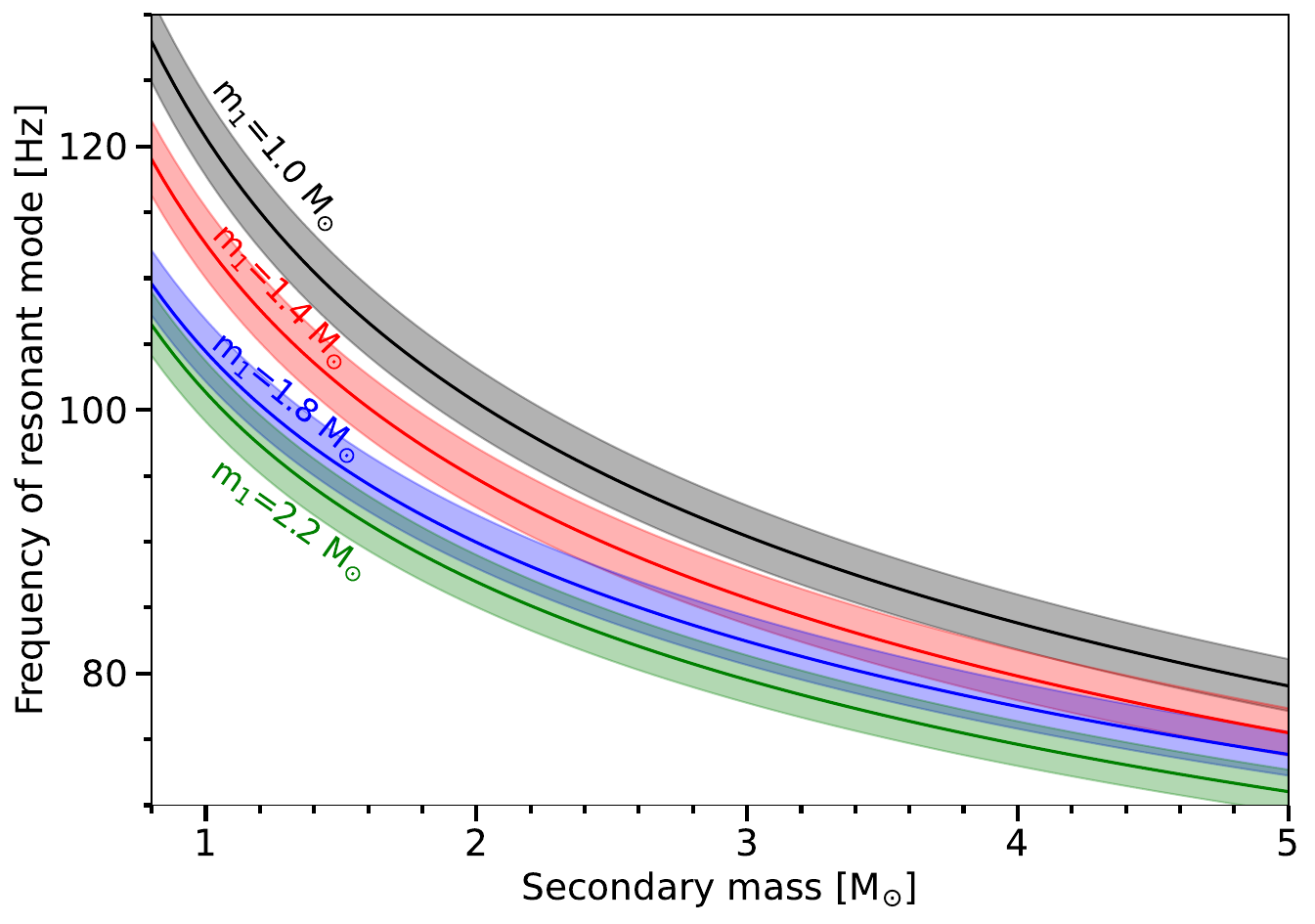}
\caption{The $i$-mode frequency which could produce a RSF with the duration of GRB 230307A's precursor, showing how it varies with the mass of one object in the binary when assuming one of several different values for the mass of the other object. The solid lines are for a duration of $0.4$ s, and the shaded regions vary this by the minimum variability timescale ($\pm17$ ms). Note that this is approximate, owing to simplifications made in the calculation of the resonance duration.}
\label{fig:duration_constraints}
\end{figure}

We fix the mass of one object in the binary and calculate the inspiral for several different secondary masses and $i$-mode frequencies, obtaining the duration of the resonance window for each case. By fitting these durations as a function of the secondary object's mass and the $i$-mode frequency we obtain an approximate relationship for the duration of the $i$-mode's resonance. We repeat this for several different fixed masses. Using the resulting relationships, in Figure~\ref{fig:duration_constraints} we show the approximate $i$-mode frequency required to produce a RSF with the duration of GRB 230307A's precursor as a function of the mass of one object in the binary, assuming four different values for the mass of the other object. Here we use the minimum variability timescale as an uncertainty in the relationship between flare duration and the duration of the resonance, as in the simplified colliding fireball shell model this timescale is the time taken for shocks to cross the fireballs shells, which is also the time by which the flare extends beyond the resonance. We find that an $i$-mode frequency of $\sim 100\text{ Hz}$ is required to produce a RSF with a $0.4\text{ s}$ duration, which is on the same scale as the values calculated using several different NS models \citep{Tsang2012}. 

Measuring the $i$-mode frequency directly would allow for strong constraints to be placed on the nuclear physics parameters that determine the nuclear matter equation of state \citep{Neill2021, Neill2023}. This direct measurement requires a EM/GW multimessenger detection of an RSF, however, we have shown above that a precursor duration can provide a combined $i$-mode frequency/chirp mass contraint, that may help to inform future detections. 

\subsection{Other pre-merger models}

Other models for short GRB precursors commonly rely on magnetospheric interactions between a NS and a BH or another NS. A BH binary partner could generate a non-thermal flare as it orbits and spins inside the magnetosphere of the NS, accelerating plasma along the field lines and thus producing gamma-rays \citep{McWilliams2011}. This emission could reach the luminosity of the precursor to GRB 230307A for a NS with dipole magnetic field strength $\gtrsim 10^{15}$~G and a relatively low mass BH. However, such a flare would appear as a very short and sharp peak. This profile is completely different compared to the one observed for GRB 230307A.

A precursor signal could be also produced in binary NS systems, when two strongly magnetized NSs in a close binary are connected by a flux tube. If the NSs have different spins, or their spins are misaligned with their magnetic fields, this flux tube will become twisted. This transfers energy from the NS's spins into the magnetic field, building the pressure needed to produce a bright non-thermal flare \citep{Most2020,Beloborodov2021}. The luminosity of this flare will be strongly dependent on the NS magnetic field strength, and also requires a dipole magnetic field strength $\gtrsim 10^{15}$~G to reach the luminosity of GRB~230307A. Also this mechanism requires that both NSs be strongly magnetized, which as previously mentioned is difficult to explain.

A common theme for many of these models is the requirement that at least one NS in the binary be very strongly magnetized (surface dipole field strength $\gtrsim 10^{15}\text{ G}$) in order to reach a luminosity $\gtrsim10^{50}$ erg/s. Thus, if GRB~230307A is originated in a binary NS merger, the energetic of this precursor strongly suggest that the NS can maintain a strong magnetic field over the duration of the inspiral phase \citep[e.g. ][]{Beniamini2023}, indicating that it is preserved by some specific mechanism \cite[e.g. it could be frozen into the superfluid NS core. See ][]{Ho2017}.

\subsection{Post-merger models}

In the post-merger phase, the precursor emission could be related to either the remnant magnetar \citep{Lipunov2008}, the evolution of the relativistic fireball \citep{MeszarosReees2000,Daigne2002}, or the interaction of the jet with the merger ejecta \citep{Troja2010,Nagakura2014,LiYu2016,Nakar2020}. 
Among these possible models, both the fireball and the jet scenarios fail to reproduce the observed properties. 
The fireball is initially optically thick but  emission from its photosphere becomes visible as soon as the expanding material becomes transparent \citep{MeszarosReees2000,Daigne2002}. However, this would produce a thermal spectral shape, at odds with observations. 
A quasi-thermal spectrum would also be expected from the jet breaking out of the ejecta cloud. Although it has been argued that the shock break-out spectrum could differ significantly from a thermal shape \citep{Bromberg2018,Nakar2020}, this model remains challenged by the high luminosity and short variability timescale of the candidate precursor. 

Models related to the central engine offer many more degrees of freedom, and thus can more easily account for a broad range of behaviors. 
If the merger remnant is a rapidly spinning, highly magnetized NS, its rotational energy ($\approx$10$^{52}$ erg) can easily account for luminosities of the order of 10$^{50}$ erg\,s$^{-1}$.
This energy is extracted by some MHD processes and then released into a high-entropy fireball. 
The baryon-rich environment of a proto-NS is considered less efficient in accelerating the fireball
to high Lorentz factors \citep{Ciolfi17}, and this might explain the softer spectrum of the initial pulse. 
Once the path is cleaned by the precursor shell, 
subsequent fireball shells could attain much higher velocities and produce the hard gamma-ray spectrum of the main episode. These MHD winds could be halted by the infalling material onto the central NS, causing the 0.4~s delay between the soft-energy precursor and the hard-energy main emission. 
Continued accretion of matter could cause the NS to collapse into a BH \citep{WangMeszaros2007}, which would then power the main burst and its temporally extended emission. In this scenario, the main open question remains the nature of the power mechanism, since the typical lifetime of the accretion disk, $t_{visc} \approx$0.2~$\left(0.1/\alpha\right)$ s where $\alpha$ is the dimensionless viscosity parameter \citep{ShakuraSunyaev1973},  is much shorter than the GRB duration.

Alternatively, if the NS remains stable, the GRB might be powered by its spin-down luminosity \citep[e.g. ][]{ZhangMeszaros2001, Troja2007, Rowlinson2010}. 
This depends markedly on the EoS, which determines the NS maximum mass, from 
$\approx$2.05\,M$_{\odot}$ for soft EoS \citep{Douchin01}
to $\approx$2.7\,M$_{\odot}$ for stiff EoS \citep{Glendenning91}, 
and its typical lifetime $t_{\rm\,col}$. 
To reproduce the average luminosity of the temporally extended high-energy emission ($\sim$6.5$\times$10$^{50}$ erg s$^{-1}$)
and its duration ($\sim$40 s),
the newly formed NS should have a high poloidal magnetic field, $B_p\,\sim\,6\times10^{15}\,\xi^{0.5}$ G, and a short initial spin period, $P_0\,\sim\,
1\,\xi^{0.5}$ ms, where $\xi$ is the 
gamma-ray conversion efficiency. 
As the derived period is already close to the break-up limit \citep{Lattimer07}, then most of the spin-down energy is efficiently converted into electromagnetic radiation ($\xi\,\gtrsim$50\%). 
Additionally, the collapse time must be
$t_{\rm\,col}\,\gtrsim$40\,s. 
These electromagnetic constraints on 
$B_{p}$, $P_{0}$, and $t_{\rm\,col}$, when 
combined with the GW measurements of the chirp mass,  could place tight constraints on the EoS of dense matter \citep{Lasky14}.

\section{Summary}
\label{sec:summary}

The prompt emission of GRB~230307A starts with a brief and luminous pulse of high-energy radiation, which 
we interpret as a precursor. 
This signal is characterized by a short minimum variability timescale (approximately 17 ms), a soft non-thermal spectrum peaking around 200 keV, and a negligible spectral lag.
This combination of soft spectrum and short variability deviates from the general trend of prompt GRB emission, supporting  our hypothesis of a precursor signal powered by a different mechanism.

We explore a wide range of precursor models to explain the high luminosity ($\sim$3.6$\times$10$^{50}$ erg s$^{-1}$ at 291 Mpc) and duration of this candidate precursor. Although our results are affected by the uncertainty in the GRB distance scale, resonant shattering flares occurring before the merger are a viable option. 
This model requires the progenitor NS to retain a high magnetic field ($\gtrsim 10^{15}\text{ G}$) within its core. 

Alternatively, high luminosity precursors
could be the product of a rapidly rotating millisecond pulsar
formed after the merger.  
This model naturally accounts for the long duration of the gamma-ray emission, and favors stiff EoS for the formation of stable supramassive NSs.  

A clear discriminant between these two scenarios (pre-merger and post-merger precursors) would be the simultaneous detection of gravitational waves, setting the exact time of the merger before or after the high-energy precursor.  
The upcoming LIGO/Virgo/Kagra observing run  O4 
holds promise for the possible detection of binary mergers associated with long GRBs. The enhanced sensitivity will allow to observe grativational wave signal from NS mergers out to $\sim$350 Mpc \citep{Petrov2022} testing possible association with high energy signals originated at the same distance scale of GRB~230307A or GRB~211211A. 
Such detections would finally dispel any doubt about the nature of the signals and the observations of a precursor could be used to constrain the properties of the merging neutron stars like the properties of the crust tidal resonant shattering, the magnetic field and, ultimately, the equation of state of dense matter.

\newpage

\section*{ACKNOWLEDGEMENTS}

This research has made use of data and software provided by the High Energy Astrophysics Science Archive Research Center (HEASARC), which is a service of the Astrophysics Science Division at NASA/GSFC and the High Energy Astrophysics Division of the Smithsonian Astrophysical Observatory. 

The material is based upon work supported by NASA under award
number 80NSSC22K1516.
E.T. and Y.-H.Y. were supported by the European Research Council through the Consolidator grant BHianca (grant agreement ID 101002761).
Part of this work was carried out at the Aspen Center for Physics, which is supported by National Science Foundation grant PHY-2210452.

D.T. and D.N.'s research was supported by UK Science and Technology Facilities Council grant ST/X001067/1 and Royal Society research grant RGS /\ R1 /\ 231499





\bibliography{sample631}{}

\begin{thebibliography}{}
\expandafter\ifx\csname natexlab\endcsname\relax\def\natexlab#1{#1}\fi
\providecommand{\url}[1]{\href{#1}{#1}}
\providecommand{\dodoi}[1]{doi:~\href{http://doi.org/#1}{\nolinkurl{#1}}}
\providecommand{\doeprint}[1]{\href{http://ascl.net/#1}{\nolinkurl{http://ascl.net/#1}}}
\providecommand{\doarXiv}[1]{\href{https://arxiv.org/abs/#1}{\nolinkurl{https://arxiv.org/abs/#1}}}

\bibitem[{{Band} {et~al.}(1993){Band}, {Matteson}, {Ford}, {Schaefer},
  {Palmer}, {Teegarden}, {Cline}, {Briggs}, {Paciesas}, {Pendleton}, {Fishman},
  {Kouveliotou}, {Meegan}, {Wilson}, \& {Lestrade}}]{Band1993}
{Band}, D., {Matteson}, J., {Ford}, L., {et~al.} 1993, \apj, 413, 281,
  \dodoi{10.1086/172995}

\bibitem[{{Beloborodov}(2021)}]{Beloborodov2021}
{Beloborodov}, A.~M. 2021, \apj, 921, 92, \dodoi{10.3847/1538-4357/ac17e7}

\bibitem[{{Beniamini} {et~al.}(2023){Beniamini}, {Wadiasingh}, {Hare},
  {Rajwade}, {Younes}, \& {van der Horst}}]{Beniamini2023}
{Beniamini}, P., {Wadiasingh}, Z., {Hare}, J., {et~al.} 2023, \mnras, 520,
  1872, \dodoi{10.1093/mnras/stad208}

\bibitem[{{Bromberg} {et~al.}(2018){Bromberg}, {Tchekhovskoy}, {Gottlieb},
  {Nakar}, \& {Piran}}]{Bromberg2018}
{Bromberg}, O., {Tchekhovskoy}, A., {Gottlieb}, O., {Nakar}, E., \& {Piran}, T.
  2018, \mnras, 475, 2971, \dodoi{10.1093/mnras/stx3316}

\bibitem[{{Burlon} {et~al.}(2009){Burlon}, {Ghirlanda}, {Ghisellini},
  {Greiner}, \& {Celotti}}]{Burlon09}
{Burlon}, D., {Ghirlanda}, G., {Ghisellini}, G., {Greiner}, J., \& {Celotti},
  A. 2009, \aap, 505, 569, \dodoi{10.1051/0004-6361/200912662}

\bibitem[{{Burns} {et~al.}(2023){Burns}, {Goldstein}, {Lesage}, {Dalessi}, \&
  {Fermi-GBM Team.}}]{Burns2023GCN}
{Burns}, E., {Goldstein}, A., {Lesage}, S., {Dalessi}, S., \& {Fermi-GBM Team.}
  2023, GRB Coordinates Network, 33414, 1

\bibitem[{{Camisasca} {et~al.}(2023){Camisasca}, {Guidorzi}, {Bulla}, {Amati},
  {Rossi}, {Stratta}, \& {Singh}}]{Camisasca2023GCN}
{Camisasca}, A.~E., {Guidorzi}, C., {Bulla}, M., {et~al.} 2023, GRB Coordinates
  Network, 33577, 1

\bibitem[{{Casentini} {et~al.}(2023){Casentini}, {Tavani}, {Pittori},
  {Lucarelli}, {Verrecchia}, {Argan}, {Cardillo}, {Evangelista}, {Foffano},
  {Piano}, {Addis}, {Baroncelli}, {Bulgarelli}, {di Piano}, {Fioretti},
  {Panebianco}, {Parmiggiani}, {Longo}, {Romani}, {Marisaldi}, {Pilia},
  {Trois}, {Donnarumma}, {Menegoni}, {Ursi}, {Giuliani}, {Tempesta}, \& {Agile
  Team}}]{GCN33412}
{Casentini}, C., {Tavani}, M., {Pittori}, C., {et~al.} 2023, GRB Coordinates
  Network, 33412, 1

\bibitem[{{Ciolfi} {et~al.}(2017){Ciolfi}, {Kastaun}, {Giacomazzo}, {Endrizzi},
  {Siegel}, \& {Perna}}]{Ciolfi17}
{Ciolfi}, R., {Kastaun}, W., {Giacomazzo}, B., {et~al.} 2017, \prd, 95, 063016,
  \dodoi{10.1103/PhysRevD.95.063016}

\bibitem[{{Cooper} {et~al.}(2023){Cooper}, {Gupta}, {Wadiasingh}, {Wijers},
  {Boersma}, {Andreoni}, {Rowlinson}, \& {Gourdji}}]{Cooper2023}
{Cooper}, A.~J., {Gupta}, O., {Wadiasingh}, Z., {et~al.} 2023, \mnras, 519,
  3923, \dodoi{10.1093/mnras/stac3580}

\bibitem[{{Coppin} {et~al.}(2020){Coppin}, {de Vries}, \& {van
  Eijndhoven}}]{Coppin2020}
{Coppin}, P., {de Vries}, K.~D., \& {van Eijndhoven}, N. 2020, \prd, 102,
  103014, \dodoi{10.1103/PhysRevD.102.103014}

\bibitem[{{Daigne} \& {Mochkovitch}(2002)}]{Daigne2002}
{Daigne}, F., \& {Mochkovitch}, R. 2002, \mnras, 336, 1271,
  \dodoi{10.1046/j.1365-8711.2002.05875.x}

\bibitem[{{Dalessi} \& {Fermi GBM Team}(2023)}]{GCN33551}
{Dalessi}, S., \& {Fermi GBM Team}. 2023, GRB Coordinates Network, 33551, 1

\bibitem[{{Dalessi} {et~al.}(2023){Dalessi}, {Roberts}, {Meegan}, \& {Fermi GBM
  Team}}]{GCN33411}
{Dalessi}, S., {Roberts}, O.~J., {Meegan}, C., \& {Fermi GBM Team}. 2023, GRB
  Coordinates Network, 33411, 1

\bibitem[{{Douchin} \& {Haensel}(2001)}]{Douchin01}
{Douchin}, F., \& {Haensel}, P. 2001, \aap, 380, 151,
  \dodoi{10.1051/0004-6361:20011402}

\bibitem[{{Fenimore} {et~al.}(1995){Fenimore}, {in 't Zand}, {Norris},
  {Bonnell}, \& {Nemiroff}}]{Fenimore1995}
{Fenimore}, E.~E., {in 't Zand}, J.~J.~M., {Norris}, J.~P., {Bonnell}, J.~T.,
  \& {Nemiroff}, R.~J. 1995, \apjl, 448, L101, \dodoi{10.1086/309603}

\bibitem[{{Fermi GBM Team}(2023)}]{2023GCN.33405....1F}
{Fermi GBM Team}. 2023, GRB Coordinates Network, 33405, 1

\bibitem[{{Fermi Science Support Development Team}(2019)}]{2019ascl.soft05011F}
{Fermi Science Support Development Team}. 2019, {Fermitools: Fermi Science
  Tools}, Astrophysics Source Code Library, record ascl:1905.011.
\newblock \doeprint{1905.011}

\bibitem[{{Gamma-ray astronomy Group}(2014)}]{2014ascl.soft09011G}
{Gamma-ray astronomy Group}, U. o. A.~H. 2014, {rmfit: Forward-folding spectral
  analysis software}, Astrophysics Source Code Library, record ascl:1409.011.
\newblock \doeprint{1409.011}

\bibitem[{{Gehrels} {et~al.}(2006){Gehrels}, {Norris}, {Barthelmy}, {Granot},
  {Kaneko}, {Kouveliotou}, {Markwardt}, {M{\'e}sz{\'a}ros}, {Nakar}, {Nousek},
  {O'Brien}, {Page}, {Palmer}, {Parsons}, {Roming}, {Sakamoto}, {Sarazin},
  {Schady}, {Stamatikos}, \& {Woosley}}]{Gehrels06}
{Gehrels}, N., {Norris}, J.~P., {Barthelmy}, S.~D., {et~al.} 2006, \nat, 444,
  1044, \dodoi{10.1038/nature05376}

\bibitem[{{Glendenning} \& {Moszkowski}(1991)}]{Glendenning91}
{Glendenning}, N.~K., \& {Moszkowski}, S.~A. 1991, \prl, 67, 2414,
  \dodoi{10.1103/PhysRevLett.67.2414}

\bibitem[{{Golkhou} \& {Butler}(2014)}]{Golkhou2014}
{Golkhou}, V.~Z., \& {Butler}, N.~R. 2014, \apj, 787, 90,
  \dodoi{10.1088/0004-637X/787/1/90}

\bibitem[{{Gourgouliatos} {et~al.}(2016){Gourgouliatos}, {Wood}, \&
  {Hollerbach}}]{Gourgouliatos2016}
{Gourgouliatos}, K.~N., {Wood}, T.~S., \& {Hollerbach}, R. 2016, Proceedings of
  the National Academy of Science, 113, 3944, \dodoi{10.1073/pnas.1522363113}

\bibitem[{{Gruber} {et~al.}(2014){Gruber}, {Goldstein}, {Weller von Ahlefeld},
  {Narayana Bhat}, {Bissaldi}, {Briggs}, {Byrne}, {Cleveland}, {Connaughton},
  {Diehl}, {Fishman}, {Fitzpatrick}, {Foley}, {Gibby}, {Giles}, {Greiner},
  {Guiriec}, {van der Horst}, {von Kienlin}, {Kouveliotou}, {Layden}, {Lin},
  {Meegan}, {McGlynn}, {Paciesas}, {Pelassa}, {Preece}, {Rau}, {Wilson-Hodge},
  {Xiong}, {Younes}, \& {Yu}}]{Gruber2014}
{Gruber}, D., {Goldstein}, A., {Weller von Ahlefeld}, V., {et~al.} 2014, \apjs,
  211, 12, \dodoi{10.1088/0067-0049/211/1/12}

\bibitem[{{Ho} {et~al.}(2017){Ho}, {Andersson}, \& {Graber}}]{Ho2017}
{Ho}, W. C.~G., {Andersson}, N., \& {Graber}, V. 2017, \prc, 96, 065801,
  \dodoi{10.1103/PhysRevC.96.065801}

\bibitem[{{Kouveliotou} {et~al.}(1993){Kouveliotou}, {Meegan}, {Fishman},
  {Bhat}, {Briggs}, {Koshut}, {Paciesas}, \& {Pendleton}}]{Kouveliotou1993}
{Kouveliotou}, C., {Meegan}, C.~A., {Fishman}, G.~J., {et~al.} 1993, \apjl,
  413, L101, \dodoi{10.1086/186969}

\bibitem[{{Kozyrev} {et~al.}(2023){Kozyrev}, {Golovin}, {Litvak}, {Mitrofanov},
  {Sanin}, {Hend/Mars Odyssey Team}, {Svinkin}, {Lysenko}, {Ridnaia}, {Ipn},
  {Goldstein}, {Briggs}, {Wilson-Hodge}, {Burns}, {Fermi Gbm Team}, {Bozzo},
  {Ferrigno}, {INTEGRAL SPI-ACS Grb Team}, {Barthelmy}, {Cummings}, {Krimm},
  {Palmer}, {Tohuvavohu}, {Swift-Bat Team}, {Boynton}, {Fellows}, {Harshman},
  {Enos}, {Starr}, {Gardner}, \& {Grs-Odyssey Grb Team}}]{GCN33413}
{Kozyrev}, A.~S., {Golovin}, D.~V., {Litvak}, M.~L., {et~al.} 2023, GRB
  Coordinates Network, 33413, 1

\bibitem[{{Lai}(1994)}]{Lai1994}
{Lai}, D. 1994, \mnras, 270, 611, \dodoi{10.1093/mnras/270.3.611}

\bibitem[{{Lasky} {et~al.}(2014){Lasky}, {Haskell}, {Ravi}, {Howell}, \&
  {Coward}}]{Lasky14}
{Lasky}, P.~D., {Haskell}, B., {Ravi}, V., {Howell}, E.~J., \& {Coward}, D.~M.
  2014, \prd, 89, 047302, \dodoi{10.1103/PhysRevD.89.047302}

\bibitem[{{Lattimer} \& {Prakash}(2007)}]{Lattimer07}
{Lattimer}, J.~M., \& {Prakash}, M. 2007, \physrep, 442, 109,
  \dodoi{10.1016/j.physrep.2007.02.003}

\bibitem[{{Lazzati}(2005)}]{Lazzati05}
{Lazzati}, D. 2005, \mnras, 357, 722, \dodoi{10.1111/j.1365-2966.2005.08687.x}

\bibitem[{{Li} \& {Yu}(2016)}]{LiYu2016}
{Li}, S.-Z., \& {Yu}, Y.-W. 2016, \apj, 819, 120,
  \dodoi{10.3847/0004-637X/819/2/120}

\bibitem[{{Li} {et~al.}(2021){Li}, {Zhang}, {Zhang}, \& {Zhen}}]{Li2021}
{Li}, X.~J., {Zhang}, Z.~B., {Zhang}, X.~L., \& {Zhen}, H.~Y. 2021, \apjs, 252,
  16, \dodoi{10.3847/1538-4365/abd3fd}

\bibitem[{{Lipunov} {et~al.}(2008){Lipunov}, {Kornilov}, {Gorbovskoy},
  {Krylov}, {Tyurina}, {Kuvshinov}, {Belinski}, {Gritsyk}, {Sankovich}, \&
  {Vladimirov}}]{Lipunov2008}
{Lipunov}, V.~M., {Kornilov}, V.~G., {Gorbovskoy}, E.~S., {et~al.} 2008,
  Astronomy Letters, 34, 145, \dodoi{10.1007/s11443-008-3002-5}

\bibitem[{{McWilliams} \& {Levin}(2011)}]{McWilliams2011}
{McWilliams}, S.~T., \& {Levin}, J. 2011, \apj, 742, 90,
  \dodoi{10.1088/0004-637X/742/2/90}

\bibitem[{{M{\'e}sz{\'a}ros} \& {Rees}(2000)}]{MeszarosReees2000}
{M{\'e}sz{\'a}ros}, P., \& {Rees}, M.~J. 2000, \apj, 530, 292,
  \dodoi{10.1086/308371}

\bibitem[{{Most} \& {Philippov}(2020)}]{Most2020}
{Most}, E.~R., \& {Philippov}, A.~A. 2020, \apjl, 893, L6,
  \dodoi{10.3847/2041-8213/ab8196}

\bibitem[{{Nagakura} {et~al.}(2014){Nagakura}, {Hotokezaka}, {Sekiguchi},
  {Shibata}, \& {Ioka}}]{Nagakura2014}
{Nagakura}, H., {Hotokezaka}, K., {Sekiguchi}, Y., {Shibata}, M., \& {Ioka}, K.
  2014, \apjl, 784, L28, \dodoi{10.1088/2041-8205/784/2/L28}

\bibitem[{{Nakar}(2020)}]{Nakar2020}
{Nakar}, E. 2020, \physrep, 886, 1, \dodoi{10.1016/j.physrep.2020.08.008}

\bibitem[{{Nasa High Energy Astrophysics Science Archive Research Center
  (Heasarc)}(2014)}]{2014ascl.soft08004N}
{Nasa High Energy Astrophysics Science Archive Research Center (Heasarc)}.
  2014, {HEAsoft: Unified Release of FTOOLS and XANADU}, Astrophysics Source
  Code Library, record ascl:1408.004.
\newblock \doeprint{1408.004}

\bibitem[{{Neill} {et~al.}(2021){Neill}, {Newton}, \& {Tsang}}]{Neill2021}
{Neill}, D., {Newton}, W.~G., \& {Tsang}, D. 2021, \mnras, 504, 1129,
  \dodoi{10.1093/mnras/stab764}

\bibitem[{{Neill} {et~al.}(2023){Neill}, {Preston}, {Newton}, \&
  {Tsang}}]{Neill2023}
{Neill}, D., {Preston}, R., {Newton}, W.~G., \& {Tsang}, D. 2023, \prl, 130,
  112701, \dodoi{10.1103/PhysRevLett.130.112701}

\bibitem[{{Neill} {et~al.}(2022){Neill}, {Tsang}, {van Eerten}, {Ryan}, \&
  {Newton}}]{Neill2022}
{Neill}, D., {Tsang}, D., {van Eerten}, H., {Ryan}, G., \& {Newton}, W.~G.
  2022, \mnras, 514, 5385, \dodoi{10.1093/mnras/stac1645}

\bibitem[{{Norris} \& {Bonnell}(2006)}]{Norris06}
{Norris}, J.~P., \& {Bonnell}, J.~T. 2006, \apj, 643, 266,
  \dodoi{10.1086/502796}

\bibitem[{{O'Connor} {et~al.}(2023){O'Connor}, {Troja}, {Ryan}, {Beniamini},
  {van Eerten}, {Granot}, {Dichiara}, {Ricci}, {Lipunov}, {Gillanders}, {Gill},
  {Moss}, {Anand}, {Andreoni}, {Becerra}, {Buckley}, {Butler}, {Cenko},
  {Chasovnikov}, {Durbak}, {Francile}, {Hammerstein}, {van der Horst},
  {Kasliwal}, {Kouveliotou}, {Kutyrev}, {Lee}, {Srinivasaragavan}, {Topolev},
  {Watson}, {Yang}, \& {Zhirkov}}]{Oconnor2023}
{O'Connor}, B., {Troja}, E., {Ryan}, G., {et~al.} 2023, arXiv e-prints,
  arXiv:2302.07906, \dodoi{10.48550/arXiv.2302.07906}

\bibitem[{{Palenzuela} {et~al.}(2013){Palenzuela}, {Lehner}, {Ponce},
  {Liebling}, {Anderson}, {Neilsen}, \& {Motl}}]{Palenzuela13}
{Palenzuela}, C., {Lehner}, L., {Ponce}, M., {et~al.} 2013, \prl, 111, 061105,
  \dodoi{10.1103/PhysRevLett.111.061105}

\bibitem[{{Petrov} {et~al.}(2022){Petrov}, {Singer}, {Coughlin}, {Kumar},
  {Almualla}, {Anand}, {Bulla}, {Dietrich}, {Foucart}, \&
  {Guessoum}}]{Petrov2022}
{Petrov}, P., {Singer}, L.~P., {Coughlin}, M.~W., {et~al.} 2022, \apj, 924, 54,
  \dodoi{10.3847/1538-4357/ac366d}

\bibitem[{{Planck Collaboration} {et~al.}(2020){Planck Collaboration},
  {Aghanim}, {Akrami}, {Ashdown}, {Aumont}, {Baccigalupi}, {Ballardini},
  {Banday}, {Barreiro}, {Bartolo}, {Basak}, {Battye}, {Benabed}, {Bernard},
  {Bersanelli}, {Bielewicz}, {Bock}, {Bond}, {Borrill}, {Bouchet}, {Boulanger},
  {Bucher}, {Burigana}, {Butler}, {Calabrese}, {Cardoso}, {Carron},
  {Challinor}, {Chiang}, {Chluba}, {Colombo}, {Combet}, {Contreras}, {Crill},
  {Cuttaia}, {de Bernardis}, {de Zotti}, {Delabrouille}, {Delouis}, {Di
  Valentino}, {Diego}, {Dor{\'e}}, {Douspis}, {Ducout}, {Dupac}, {Dusini},
  {Efstathiou}, {Elsner}, {En{\ss}lin}, {Eriksen}, {Fantaye}, {Farhang},
  {Fergusson}, {Fernandez-Cobos}, {Finelli}, {Forastieri}, {Frailis},
  {Fraisse}, {Franceschi}, {Frolov}, {Galeotta}, {Galli}, {Ganga},
  {G{\'e}nova-Santos}, {Gerbino}, {Ghosh}, {Gonz{\'a}lez-Nuevo}, {G{\'o}rski},
  {Gratton}, {Gruppuso}, {Gudmundsson}, {Hamann}, {Handley}, {Hansen},
  {Herranz}, {Hildebrandt}, {Hivon}, {Huang}, {Jaffe}, {Jones}, {Karakci},
  {Keih{\"a}nen}, {Keskitalo}, {Kiiveri}, {Kim}, {Kisner}, {Knox},
  {Krachmalnicoff}, {Kunz}, {Kurki-Suonio}, {Lagache}, {Lamarre}, {Lasenby},
  {Lattanzi}, {Lawrence}, {Le Jeune}, {Lemos}, {Lesgourgues}, {Levrier},
  {Lewis}, {Liguori}, {Lilje}, {Lilley}, {Lindholm}, {L{\'o}pez-Caniego},
  {Lubin}, {Ma}, {Mac{\'\i}as-P{\'e}rez}, {Maggio}, {Maino}, {Mandolesi},
  {Mangilli}, {Marcos-Caballero}, {Maris}, {Martin}, {Martinelli},
  {Mart{\'\i}nez-Gonz{\'a}lez}, {Matarrese}, {Mauri}, {McEwen}, {Meinhold},
  {Melchiorri}, {Mennella}, {Migliaccio}, {Millea}, {Mitra},
  {Miville-Desch{\^e}nes}, {Molinari}, {Montier}, {Morgante}, {Moss}, {Natoli},
  {N{\o}rgaard-Nielsen}, {Pagano}, {Paoletti}, {Partridge}, {Patanchon},
  {Peiris}, {Perrotta}, {Pettorino}, {Piacentini}, {Polastri}, {Polenta},
  {Puget}, {Rachen}, {Reinecke}, {Remazeilles}, {Renzi}, {Rocha}, {Rosset},
  {Roudier}, {Rubi{\~n}o-Mart{\'\i}n}, {Ruiz-Granados}, {Salvati}, {Sandri},
  {Savelainen}, {Scott}, {Shellard}, {Sirignano}, {Sirri}, {Spencer},
  {Sunyaev}, {Suur-Uski}, {Tauber}, {Tavagnacco}, {Tenti}, {Toffolatti},
  {Tomasi}, {Trombetti}, {Valenziano}, {Valiviita}, {Van Tent}, {Vibert},
  {Vielva}, {Villa}, {Vittorio}, {Wandelt}, {Wehus}, {White}, {White},
  {Zacchei}, \& {Zonca}}]{Planck2018}
{Planck Collaboration}, {Aghanim}, N., {Akrami}, Y., {et~al.} 2020, \aap, 641,
  A6, \dodoi{10.1051/0004-6361/201833910}

\bibitem[{{Ramirez-Ruiz} \& {Merloni}(2001)}]{Ramirez-Ruiz2001}
{Ramirez-Ruiz}, E., \& {Merloni}, A. 2001, \mnras, 320, L25,
  \dodoi{10.1046/j.1365-8711.2001.04130.x}

\bibitem[{{Rastinejad} {et~al.}(2022){Rastinejad}, {Gompertz}, {Levan}, {Fong},
  {Nicholl}, {Lamb}, {Malesani}, {Nugent}, {Oates}, {Tanvir}, {de Ugarte
  Postigo}, {Kilpatrick}, {Moore}, {Metzger}, {Ravasio}, {Rossi}, {Schroeder},
  {Jencson}, {Sand}, {Smith}, {Ag{\"u}{\'\i} Fern{\'a}ndez}, {Berger},
  {Blanchard}, {Chornock}, {Cobb}, {De Pasquale}, {Fynbo}, {Izzo}, {Kann},
  {Laskar}, {Marini}, {Paterson}, {Escorial}, {Sears}, \&
  {Th{\"o}ne}}]{Rastinejad2022}
{Rastinejad}, J.~C., {Gompertz}, B.~P., {Levan}, A.~J., {et~al.} 2022, \nat,
  612, 223, \dodoi{10.1038/s41586-022-05390-w}

\bibitem[{{Rowlinson} {et~al.}(2010){Rowlinson}, {O'Brien}, {Tanvir}, {Zhang},
  {Evans}, {Lyons}, {Levan}, {Willingale}, {Page}, {Onal}, {Burrows},
  {Beardmore}, {Ukwatta}, {Berger}, {Hjorth}, {Fruchter}, {Tunnicliffe}, {Fox},
  \& {Cucchiara}}]{Rowlinson2010}
{Rowlinson}, A., {O'Brien}, P.~T., {Tanvir}, N.~R., {et~al.} 2010, \mnras, 409,
  531, \dodoi{10.1111/j.1365-2966.2010.17354.x}

\bibitem[{{Schnittman} {et~al.}(2018){Schnittman}, {Dal Canton}, {Camp},
  {Tsang}, \& {Kelly}}]{Schnittman2018}
{Schnittman}, J.~D., {Dal Canton}, T., {Camp}, J., {Tsang}, D., \& {Kelly},
  B.~J. 2018, \apj, 853, 123, \dodoi{10.3847/1538-4357/aaa08b}

\bibitem[{{Shakura} \& {Sunyaev}(1973)}]{ShakuraSunyaev1973}
{Shakura}, N.~I., \& {Sunyaev}, R.~A. 1973, \aap, 24, 337

\bibitem[{{Sun} {et~al.}(2023){Sun}, {Wang}, {Yang}, {Zhang}, {Xiong}, {Yin},
  {Liu}, {Li}, {Xue}, {Yan}, {Zhang}, {Tan}, {Pan}, {Liu}, {Cheng}, {Zhang},
  {Hu}, {Zheng}, {An}, {Cai}, {Hu}, {Jin}, {Li}, {Li}, {Liu}, {Liu}, {Peng},
  {Song}, {Sun}, {Sun}, {Wang}, {Wen}, {Xiao}, {Yi}, {Zhang}, {Zhang}, {Zhang},
  {Zhang}, {Zhao}, {Zheng}, {Ling}, {Zhang}, {Yuan}, \& {Zhang}}]{Sun2023}
{Sun}, H., {Wang}, C.~W., {Yang}, J., {et~al.} 2023, arXiv e-prints,
  arXiv:2307.05689, \dodoi{10.48550/arXiv.2307.05689}

\bibitem[{{Suvorov} \& {Kokkotas}(2021)}]{Suvorov2020}
{Suvorov}, A.~G., \& {Kokkotas}, K.~D. 2021, \mnras, 502, 2482,
  \dodoi{10.1093/mnras/stab153}

\bibitem[{{Svinkin} {et~al.}(2023){Svinkin}, {Frederiks}, {Ulanov},
  {Tsvetkova}, {Lysenko}, {Ridnaia}, {Cline}, \& {Konus-Wind Team}}]{GCN33427}
{Svinkin}, D., {Frederiks}, D., {Ulanov}, M., {et~al.} 2023, GRB Coordinates
  Network, 33427, 1

\bibitem[{{Troja} {et~al.}(2010){Troja}, {Rosswog}, \& {Gehrels}}]{Troja2010}
{Troja}, E., {Rosswog}, S., \& {Gehrels}, N. 2010, \apj, 723, 1711,
  \dodoi{10.1088/0004-637X/723/2/1711}

\bibitem[{{Troja} {et~al.}(2007){Troja}, {Cusumano}, {O'Brien}, {Zhang},
  {Sbarufatti}, {Mangano}, {Willingale}, {Chincarini}, {Osborne}, {Marshall},
  {Burrows}, {Campana}, {Gehrels}, {Guidorzi}, {Krimm}, {La Parola}, {Liang},
  {Mineo}, {Moretti}, {Page}, {Romano}, {Tagliaferri}, {Zhang}, {Page}, \&
  {Schady}}]{Troja2007}
{Troja}, E., {Cusumano}, G., {O'Brien}, P.~T., {et~al.} 2007, \apj, 665, 599,
  \dodoi{10.1086/519450}

\bibitem[{{Troja} {et~al.}(2022){Troja}, {Fryer}, {O'Connor}, {Ryan},
  {Dichiara}, {Kumar}, {Ito}, {Gupta}, {Wollaeger}, {Norris}, {Kawai},
  {Butler}, {Aryan}, {Misra}, {Hosokawa}, {Murata}, {Niwano}, {Pandey},
  {Kutyrev}, {van Eerten}, {Chase}, {Hu}, {Caballero-Garcia}, \&
  {Castro-Tirado}}]{Troja22}
{Troja}, E., {Fryer}, C.~L., {O'Connor}, B., {et~al.} 2022, \nat, 612, 228,
  \dodoi{10.1038/s41586-022-05327-3}

\bibitem[{{Tsang}(2013)}]{Tsang2013}
{Tsang}, D. 2013, \apj, 777, 103, \dodoi{10.1088/0004-637X/777/2/103}

\bibitem[{{Tsang} {et~al.}(2012){Tsang}, {Read}, {Hinderer}, {Piro}, \&
  {Bondarescu}}]{Tsang2012}
{Tsang}, D., {Read}, J.~S., {Hinderer}, T., {Piro}, A.~L., \& {Bondarescu}, R.
  2012, \prl, 108, 011102, \dodoi{10.1103/PhysRevLett.108.011102}

\bibitem[{{Wang} {et~al.}(2020){Wang}, {Peng}, {Zou}, {Zhang}, \&
  {Zhang}}]{Wang2020}
{Wang}, J.-S., {Peng}, Z.-K., {Zou}, J.-H., {Zhang}, B.-B., \& {Zhang}, B.
  2020, \apjl, 902, L42, \dodoi{10.3847/2041-8213/abbfb8}

\bibitem[{{Wang} \& {M{\'e}sz{\'a}ros}(2007)}]{WangMeszaros2007}
{Wang}, X.-Y., \& {M{\'e}sz{\'a}ros}, P. 2007, \apj, 670, 1247,
  \dodoi{10.1086/522820}

\bibitem[{{Xiao} \& {Krucker}(2023)}]{GCN33410}
{Xiao}, H., \& {Krucker}, S. 2023, GRB Coordinates Network, 33410, 1

\bibitem[{{Xiao} {et~al.}(2022){Xiao}, {Zhang}, {Zhu}, {Xiong}, {Gao}, {Xu},
  {Zhang}, {Peng}, {Li}, {Zhang}, {Lu}, {Lin}, {Liu}, {Zhang}, {Ge}, {Tuo},
  {Xue}, {Fu}, {Liu}, {Li}, {Wang}, {Zheng}, {Wang}, {Jiang}, {Li}, {Liu},
  {Cao}, {Cai}, {Yi}, {Zhao}, {Xie}, {Li}, {Luo}, {Liao}, {Song}, {Zhang},
  {Qu}, {Liu}, {Li}, {Xu}, \& {Li}}]{Xiao22}
{Xiao}, S., {Zhang}, Y.-Q., {Zhu}, Z.-P., {et~al.} 2022, arXiv e-prints,
  arXiv:2205.02186, \dodoi{10.48550/arXiv.2205.02186}

\bibitem[{{Xiong} {et~al.}(2023){Xiong}, {Wang}, {Huang}, \& {Gecam
  Team}}]{GCN33406}
{Xiong}, S., {Wang}, C., {Huang}, Y., \& {Gecam Team}. 2023, GRB Coordinates
  Network, 33406, 1

\bibitem[{{Yang} {et~al.}(2023){Yang}, {Troja}, {O'Connor}, {Fryer}, {Im},
  {Durbak}, {Paek}, {Ricci}, {De Bom}, {Gillanders}, {Castro-Tirado}, {Peng},
  {Dichiara}, {Ryan}, {van Eerten}, {Dai}, {Chang}, {Choi}, {De}, {Hu},
  {Kilpatrick}, {Kutyrev}, {Jeong}, {Lee}, {Makler}, {Navarete}, \&
  {P{\'e}rez-Garc{\'\i}a}}]{Yuhan2023}
{Yang}, Y.-H., {Troja}, E., {O'Connor}, B., {et~al.} 2023, arXiv e-prints,
  arXiv:2308.00638, \dodoi{10.48550/arXiv.2308.00638}

\bibitem[{{Zhang} \& {M{\'e}sz{\'a}ros}(2001)}]{ZhangMeszaros2001}
{Zhang}, B., \& {M{\'e}sz{\'a}ros}, P. 2001, \apjl, 552, L35,
  \dodoi{10.1086/320255}

\bibitem[{{Zhong} {et~al.}(2019){Zhong}, {Dai}, {Cheng}, {Lan}, \&
  {Zhang}}]{Zhong2019}
{Zhong}, S.-Q., {Dai}, Z.-G., {Cheng}, J.-G., {Lan}, L., \& {Zhang}, H.-M.
  2019, \apj, 884, 25, \dodoi{10.3847/1538-4357/ab3e48}

\end{thebibliography}
\bibliographystyle{aasjournal}

\label{lastpage}

\end{document}